\documentclass[runningheads]{svmult}

\usepackage{latexsym}
\usepackage{amssymb}
\usepackage{makeidx}   
\usepackage{graphicx}  
\usepackage{subeqnar}  
\usepackage{multicol}  
\usepackage{physprbb}  
\makeindex             

\begin{document}
\title*{Quantized Vorticity in Superfluid $^3$He-A:\protect\newline
Structure and Dynamics}
\toctitle{Quantized Vorticity in Superfluid $^3$He-A}
%
%
\titlerunning{Quantized Vorticity in Superfluid $^3$He-A}
%
\author{R. Blaauwgeers\inst{1,2}
\and V.B. Eltsov\inst{1,3}
\and M. Krusius\inst{1}
\and J. Ruohio\inst{1}
\and R. Schanen\inst{1,4}}
\authorrunning{R. Blaauwgeers et al.}
%
%
\institute{Low Temperature Lab, Helsinki University of Technology,
FIN-02015 HUT, Finland
\and Kamerlingh Onnes Lab, Leiden University, 2300 RA Leiden, The
Netherlands
\and Kapitza Institute, Kosygina 2, Moscow 117334, Russia
\and CRTBT-CNRS, BP 166, F-38042 Grenoble Cedex 09 FRA, France}

\maketitle              

\begin{abstract}
Superfluid $^3$He-A displays the largest variety in vortex
structure among the presently known coherent quantum systems. The
experimentally verified information comes mostly from NMR
measurements on the rotating fluid, from which the order-parameter
texture can often be worked out. The various vortex structures
differ in the topology of their order-parameter field, in energy,
critical velocity, and in their response to temporal variations in
the externally applied flow. They require different experimental
conditions for their creation. When the flow is applied in the
superfluid state, the structure with the lowest critical velocity
is formed. In $^3$He-A this leads to the various forms of
continuous (or singularity-free) vorticity. Which particular
structure is created depends on the externally applied conditions
and on the global order-parameter texture.
\end{abstract}

\section{Superfluid $^3$He}

The accepted textbook example of a superfluid has traditionally
been $^4$He-II. However, in many respects the $^3$He superfluids
display more ideal behaviour, both in their theoretical
description and their macroscopic properties. One remarkable
difference is the absence of remanent vorticity in most
experimental setups for superfluid $^3$He. In $^4$He-II, the
formation of vortex lines via mechanisms intrinsic to the
superfluid itself is observed only in rare cases, most notably in
the superflow through a sub-micron-size orifice where the vortex
lines are blown out of the immediate vicinity of the aperture and
are trapped on surface sites far away where the flow is small. In
$^3$He superfluids, where the vortex core radii are at least 100
times larger, intrinsic critical velocities can be measured simply
with bulk superfluid flowing past a flat wall. For this reason
rotating measurements have proven very efficient in their study,
quite unlike in $^4$He-II.

\begin{figure}[t!!!]
\begin{center}
\includegraphics[width=0.65\textwidth]{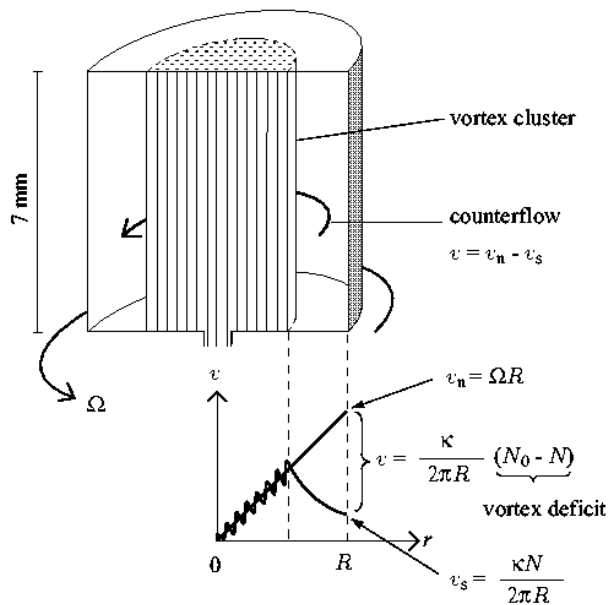}
\end{center}
\caption[RotBucket]{Vortex lines in a rotating cylindrical
container. Metastable states, which include fewer vortex lines
than the equilibrium state, consist of a central cluster with
rectilinear vortex lines and a surrounding annular region of
vortex-free counterflow. The Magnus force from the circulating
counterflow confines the lines in the cluster to the density $n_v
= 2\Omega/(\nu\kappa)$, when the cluster rotates like a solid body
with the container. If $\Omega$ is increased, the cluster
contracts, the counterflow velocity increases and ultimately
reaches the critical velocity limit $v_c$, where a new vortex is
created. If $\Omega$ is reduced, the counterflow velocity
diminishes, the cluster expands and eventually reaches the
annihilation threshold where vortices in the outermost circle of
lines are pushed to the cylindrical wall for annihilation. The
graph on the bottom illustrates how the counterflow velocity
$v(R)$ at the cylinder wall depends on the vortex-line number $N$,
when $N_0 \approx \pi R^2 n_v$ denotes their number in the
equilibrium state at the rotation velocity $\Omega$. }
\label{RotBucket}
\end{figure}

The hydrodynamics of the $^3$He superfluids abound with new
features which have only marginally been investigated. Some
experimental work has been performed on vortex tangles in the
context of quench-cooled non-equilibrium transitions from the
normal liquid to the quasi-isotropic $^3$He-B \cite{NeutronExps}
and also around vibrating wire resonators in the zero temperature
limit \cite{Fisher}. Little if any work has been reported on
turbulent flow or on vortex networks in the highly anisotropic
$^3$He-A phase \cite{Bozler}. Nevertheless, this is the phase with
the much richer variety of response to externally applied flow.
Here we describe preliminary studies in this direction, which
explain the critical flow velocity \cite{RuutuA} and the dynamic
response to rotation in the A phase \cite{Eltsov}. Unusual new
features are observed, which all can be explained by the
structural properties of the anisotropic order-parameter field,
but which also suggest that more surprises can be expected. The
critical velocity \cite{RuutuB} and the response of vortex lines
to a dynamic rotation drive \cite{Kondo} in $^3$He-B have been
measured earlier. These measurements attest to more traditional
behaviour, as can be expected for vortex lines with singular cores
in an order-parameter field with many orders of magnitude weaker
anisotropy.

Much of the existing information on new vortex structures in the
$^3$He superfluids has been derived from NMR measurements on bulk
liquid samples contained in a rotating cylinder with smooth walls.
The principle of this ``rotating bucket'' method is outlined in
Fig.~\ref{RotBucket}. Here all sensors can be placed outside the
cylinder and one can investigate the undisturbed order-parameter
field in the bulk superfluid, in the presence of different types
of quantized vorticity as well as the conditions in which they are
created.

\section{Order-parameter texture and superflow in $^3$He-A}

The beauty of $^3$He-A lies in its anisotropy: Although the
underlying material, liquid $^3$He, is isotropic, an all-pervading
anisotropy arises from the condensation into a coherent p-wave
paired state. Here the two fermion quasiparticles forming a Cooper
pair have relative angular momentum $L=1$ and total spin $S=1$
\cite{VW}. The spin structure of the condensate is characterized
by the formation of spin up-up and down-down pairs so that the
total spin $\mathbf{S}$ has zero projection on an axis, which
traditionally is denoted by the unit vector $\hat{\mathbf{d}}$
$(\perp \mathbf{S})$. The Cooper pairs also have a preferred
direction for their orbital momentum, denoted by the unit vector
$\hat\ell$, so that the projection of the net orbital momentum
$\mathbf{L}$ on the $\hat\ell$ axis is positive. Finally, also the
quasiparticle excitation spectrum is anisotropic: In momentum
space the energy gap vanishes at two opposite poles of the Fermi
surface, located on the $\hat\ell$ axis.

A number of competing interactions act in unison to produce a
smooth variation of the $\hat\ell$ and $\hat{\mathbf{d}}$ vector
fields over the container volume. These vector fields, which in
the absence of singularities are smoothly continuous, are called
the orbital and spin textures. The dipole (or spin-orbit) coupling
connects the orbital and spin textures via a free-energy density
$f_D = - g_D \, (\hat {\bf d} \cdot \hat {\bf \ell})^2$. The
magnetic anisotropy energy in an externally applied magnetic field
$\bf H$ is written as $f_H = g_H \, (\hat {\bf d} \cdot {\bf
H})^2$. These two orientational forces on $\hat{\mathbf{d}}$
balance each other at a characteristic value of the field
$\mathbf{H}$. This is called the dipole field, $H_D =
\sqrt{g_D/g_H} \sim 1$ mT, below which the dipole coupling wins
and the texture becomes {\em dipole locked}, with $\hat {\bf d}$
and $\hat {\bf \ell}$ either parallel or anti-parallel. At higher
fields, $H \gg H_D$, $\hat{\bf d}$ is forced to lie perpendicular
to $\mathbf{H}$. This is the case in NMR measurements: $\hat {\bf
d}$ is everywhere contained in the plane perpendicular to
$\textbf{H}$ and varies smoothly in this plane. This texture is
coupled to the orbital alignment such that $\hat{\bf \ell}$ is
also for the most part forced parallel to $\hat {\bf d}$.


\begin{figure}[t!!!]
\centerline{\includegraphics[width=1.0\textwidth]{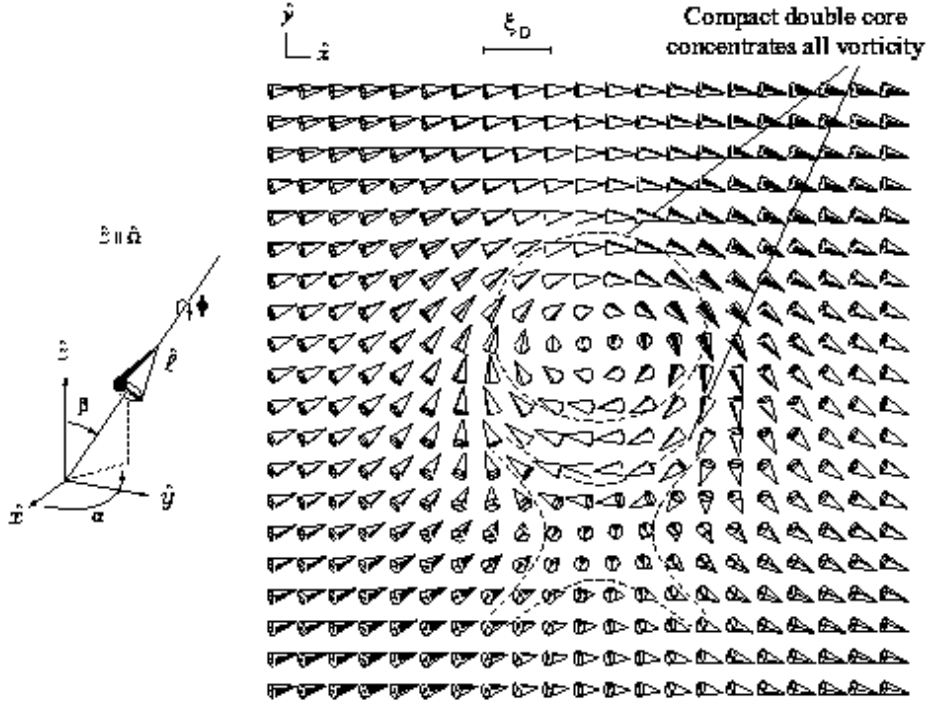}}
\caption[CUV]{Orbital $\hat {\bf \ell}$ field in the
  dipole-unlocked soft
  core of a continuous (singularity-free) vortex. The figure shows the
  orientational distribution of $\hat {\bf \ell}$ in the plane perpendicular to the
  vortex axis. It extends over a solid angle of $4\pi$,
  which corresponds to 2
  quanta of circulation. Equivalently, by following one full circle around
  the outer periphery of the figure it is seen that $\hat {\bf \ell}$
  rotates twice around its own axis, ie. the phase angle $\phi$ changes by
  $4\pi$ on circling the soft core once in the dipole-locked region far
  away where $\hat {\bf \ell}$ is uniformly oriented. This is the
  most common structure of quantized vorticity which evolves in $^3$He-A in
  a magnetic field, when rotation is started in the superfluid state. (From
  Ref.~\cite{Lexicon})}
\label{CUV}
\end{figure}

Important exceptions, when $\hat {\bf \ell}$ becomes decoupled
from the planar $\hat {\bf d}$ orientation (in the presence of a
magnetic field $H\gg H_D$), are topological defects like the {\em
dipole-unlocked} central part of a vortex line, which is also
called the ``soft vortex core'' (Fig.~\ref{CUV}). To minimize the
loss in dipole energy, the soft core has a radius on the order of
the dipolar healing length $\xi_D = \hbar/(2 m_3 v_D) \sim
10\;\mu$m, where $v_D \approx \sqrt{g_D/\rho_{s \parallel}} \sim
1$ mm/s is the order of magnitude of the so-called dipolar
velocity, ie. the flow velocity at which the orienting force on
$\hat {\bf \ell}$ from an externally applied superflow matches the
bending or gradient energy of $\hat {\bf \ell}$, which maintains
the spatial coherence of the order parameter.

The rigidity of the continuous $\hat \ell$ texture, which arises
from the anisotropy forces, from superfluid coherence, and from
the boundary conditions, explains why fluid flow in the A phase is
not always dissipative, in spite  of the gap nodes in the
quasiparticle spectrum. For most $\hat{\ell}$ textures the
velocity of the superfluid fraction $v_s$ can be nonzero, as seen
from the experimental fact that their critical velocities of
vortex formation have a finite, albeit small value. Conceptually
the single most dramatic difference from the isotropic case is the
feature that superflow does not need to remain curl-free in
$^3$He-A with gap nodes \cite{MerminHo}. The superflow velocity is
written as
\begin{equation} {\bf v}_s = {\hbar \over {2m_3}} ({\bf
\nabla} \phi - \cos{\beta} \, \nabla{\alpha}) \;,
\label{v_s}\end{equation} where $\alpha$, $\beta$, and $\phi$ are
the local azimuthal, polar, and phase angles of $\hat {\bf \ell}$.
This means that the vorticity
\begin{equation} \nabla \times {\bf v}_s = {\hbar \over
{2m_3}} \; \sin{\beta}\;(\nabla{\beta} \times \nabla{\alpha})
\label{curl_v_s}\end{equation} becomes nonvanishing in those
regions of the orbital texture where the $\hat{\bf \ell}$
orientation is not contained within one single plane. On forming
the circulation of ${\bf v}_s$ along a closed path, which
encircles such a region with inhomogeneous $\hat{\bf \ell}$
orientations,
\begin{equation} \nu \kappa  = \oint{{\bf v}_s \cdot d{\bf
r}} = {\hbar \over {2m_3}} \mathcal{S}({\hat\ell}) \;,
\label{circulation}\end{equation} one finds that the number
($\nu$) of circulation quanta ($\kappa = h/(2m_3)$) is given by
the solid angle $\mathcal{S}({\hat\ell})$ over which the $\hat
{\bf \ell}$ orientations extend within the encircled region. In
other words, the circulation is related to the topological charge
of the $\hat {\bf \ell}$ field.

In the soft core of the dipole-unlocked singularity-free vortex
(Fig.~\ref{CUV}), the orientational distribution of the $\hat {\bf
\ell}$ field covers a solid angle of $4\pi$ and $\nu =2$. This
configuration is an example of a {\em skyrmion}, which consists of
two halves, a circular and a hyperbolic Mermin-Ho vortex, which
also are known as {\em merons}. The circular half covers the
$2\pi$ orientations in the positive half sphere and the hyperbolic
those in the negative half.

\begin{figure}[t!!!]
\begin{center}
\includegraphics[width=0.7\textwidth]{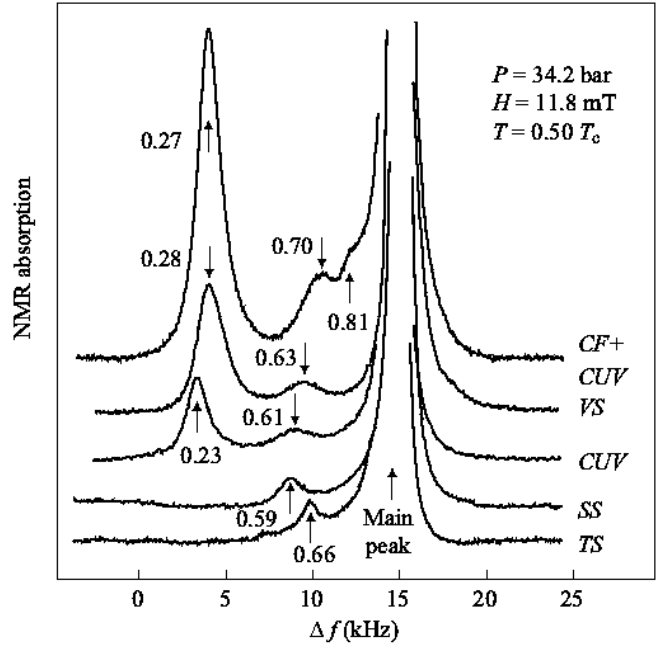}
\end{center}
\caption[NMRspectroscopy]{NMR spectroscopy of topologically stable
defects of the order-parameter field in $^3$He-A. The measured NMR
absorption is plotted as a function of the frequency shift $f-f_0$
from the Larmor value $f_0 = \gamma H_0$, where the resonance
takes place in the normal phase. The large truncated peak on the
right represents the resonance absorption of the dipole-locked
bulk superfluid at the maximum possible frequency shift. The
different satellite peaks represent: {\bf CUV} --- double-quantum
vortex lines (Fig.~\ref{CUV}) in the equilibrium rotating state at
$\Omega = 0.58$ rad/s, which in a cylinder of radius $R=2.5$ mm
corresponds to 150 vortex lines. At a temperature as low as
$0.49\,T_c$, the satellite spectrum includes both a large primary
and a small secondary peak. {\bf VS}
--- equilibrium state of the vortex sheet in the same conditions.
{\bf SS} --- soliton sheet with splay structure and with the sheet
oriented vertically parallel to the cylinder axis ($\Omega = 0$).
{\bf TS} --- soliton sheet with twist structure and oriented
transverse to the cylinder axis ($\Omega = 0.11$ rad/s, which is
below but close to $\Omega_c$). {\bf CUV + CF} --- double-quantum
vortes lines in a cluster surrounded by vortex-free counterflow
close to the critical velocity threshold ($\Omega = 2.0$ rad/s).
The two satellites with the normalized frequency shifts
$R_{\perp}^2 = 0.27$ and 0.70 are the primary and secondary
double-quantum vortex peaks while that with $R_{\perp}^2 = 0.81$
is caused by the counterflow. (In this measurement the temperature
is higher $(T=0.54 \, T_c)$ which explains the larger
$R_{\perp}^2$ values of the vortex satellites than in the CUV
spectrum.) The NMR field $\mathbf{H}$ is oriented parallel to the
rotation axis. (From Ref.~\cite{Ruutu3})} \label{NMRspectroscopy}
\end{figure}

The dipole coupling exerts an extra torque on spin precession and
gives rise to frequency shifts in NMR. Experimentally a most
valuable consequence is the fact that dipole-unlocked regions
experience a frequency shift which is different from that of the
locked bulk liquid and moreover a characteristic of the $\hat {\bf
\ell}$ texture  within the soft core of the defect. Different
structures of topological defects give rise to absorption in
satellite peaks where both the frequency shift of the peak and its
intensity are a characteristic of the defect structure
\cite{Ruutu3}. This property provides a measuring tool which
differentiates between defects and where the absorption intensity
can be calibrated to give the number of defects
(Fig.~\ref{NMRspectroscopy}).

\begin{figure}[t]
\begin{center}
\includegraphics[width=0.75\textwidth]{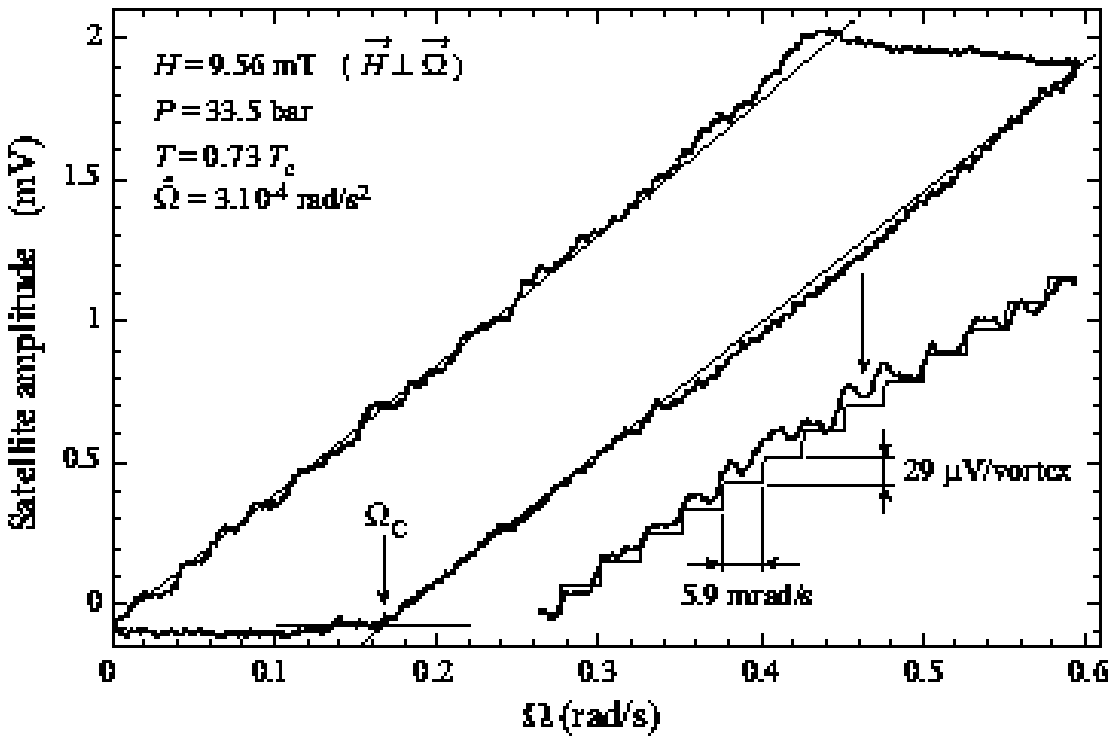}
\end{center}
\caption[Accel-Decel-Loop]{Response of double-quantum vortex lines
to a closed acceleration -- deceleration cycle in the rotation
drive $\Omega$: The peak height of the primary NMR satellite
plotted as a function of the externally applied rotation.
Initially on increasing $\Omega$ from zero, vortex-free
counterflow is created. This is known as the Landau state in
superfluids and as the Meissner state (with complete flux
expulsion) in superconductors. At $\Omega_c$ the critical
counterflow velocity, $v_c = \Omega_c R$, is reached at the
cylinder wall and the first vortex line is created. This brings
about a reduction in the flow velocity at the wall by $\Delta
\Omega \, R = \nu \kappa/(2\pi R)$. When $\Omega$ is increased
further, the process is repeated periodically and the slanting
section with the slope $dN/dt = 2\pi R^2/(\nu \kappa)$ is
recorded. This means that here the critical velocity remains
constant at $v_c = \Omega_c R = \Omega R - \nu\kappa N/(2\pi R)$.
At maximum amplification this section can be seen to mimic a
periodic signal of staircase pattern, as shown in the insert. The
upper branch of the acceleration -- deceleration cycle is measured
during decreasing $\Omega$. Here the excess vortex-free
counterflow is first reduced, until the cluster reaches the
annihilation threshold, upon which vortex lines start to
annihilate during further deceleration. At the annihilation
threshold the cluster is separated from the cylinder wall by a
counterflow annulus of minimum width: $d \approx \beta
R/(2\sqrt{\Omega})$. This is comparable to the inter-vortex
distance within the cluster: $d \approx [\nu\kappa/(2\pi
\Omega)]^{1/2}$. In practice the number of vortex lines at the
annihilation threshold \cite{Annihilation} is equal to that in the
equilibrium state: $N_0 = \pi R^2 n_v \, (1-\beta/\sqrt{\Omega})$,
where $\beta \approx 0.09$ for CUV lines. The true equilibrium
state is obtained by cooling through $T_c$ in rotation at constant
$\Omega$. Annihilation in the upper branch can be seen to be a
more random process, in which a larger number of lines may be
removed approximately simultaneously from the outermost circle of
vortex lines. (From Ref.~\cite{Blaauwgeers})}
\label{Accel-Decel-Loop}
\end{figure}

\section{Double-quantum vortex line}

The generic rotating experiment consists of an acceleration --
deceleration cycle in the rotation drive, as shown in
Fig.~\ref{Accel-Decel-Loop}. This experiment gives reproducible
results in $^3$He superfluids where remanent trapped vortex
filaments can be avoided in a container with smooth walls. In
Fig.~\ref{Accel-Decel-Loop} the NMR spectrometer has been tuned to
the frequency of the satellite from the double-quantum vortex line
and its peak height is recorded as a function of the rotation
velocity $\Omega$. The rate of change $\mid d\Omega/dt \mid \sim
10^{-4}$ rad/s$^2$ is kept as slow as possible so that dynamic
effects do not influence the result, but such that long-term
drifts neither become important. During increasing $\Omega$ it is
possible to discern from the measuring noise a stair-case-like
pattern in the satellite peak height when $\Omega\geq \Omega_c$.
The periodicity in this signal as a function of $\Omega$
calibrates the circulation associated with one vortex line:
$\Delta \Omega = \nu \kappa/(2\pi R^2)$. This provides the proof
for $\nu = 2$, namely that the vortex line created in the
externally applied flow is doubly quantized \cite{Blaauwgeers}.

\begin{figure}[t!!!]
\begin{center}
\includegraphics[width=.7\textwidth]{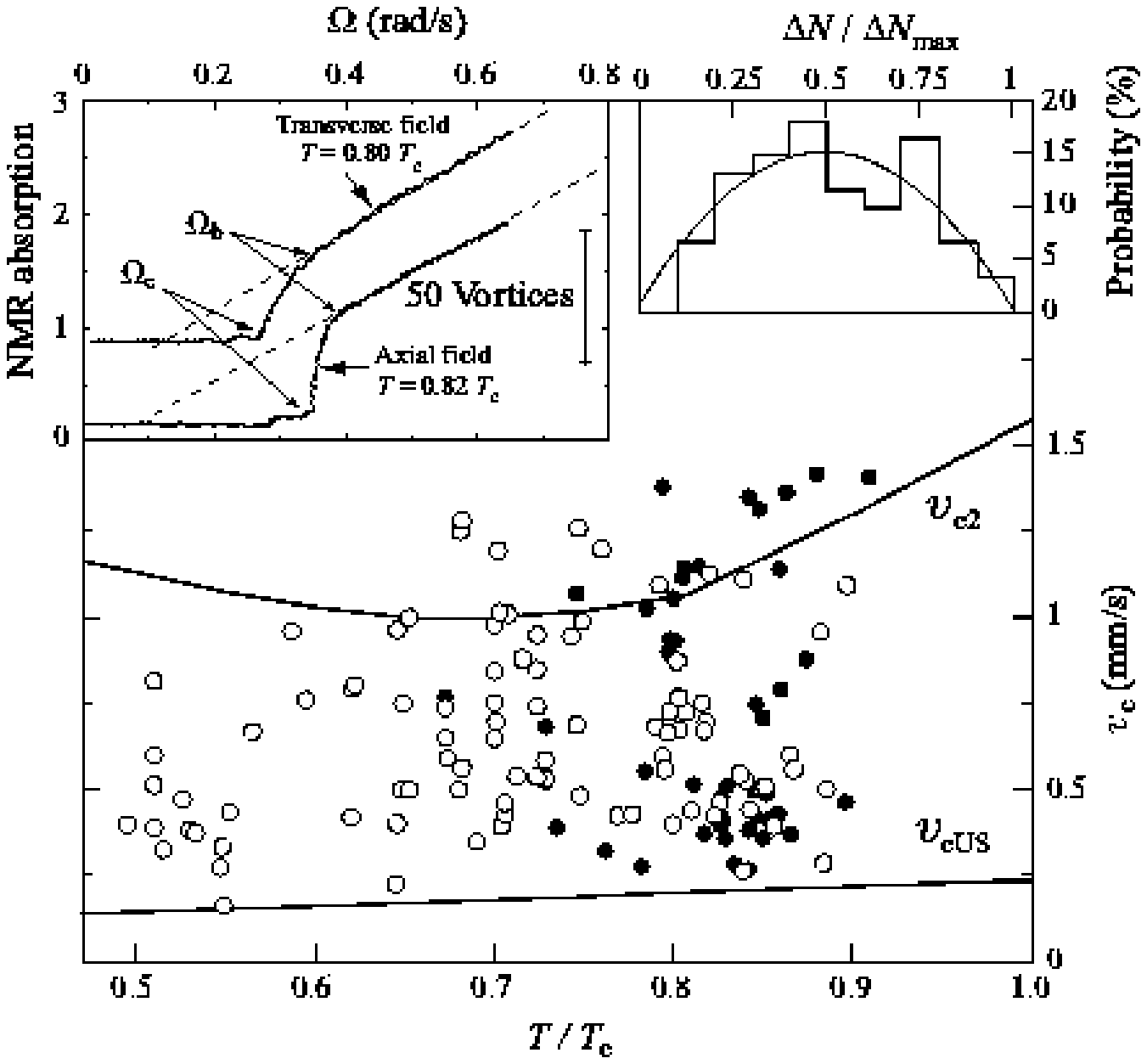}
\end{center}
\caption[CritVel-TextTransition]{Critical velocity for the
formation of the double-quantum vortex. Vortex formation can
proceed as a regular periodic process, as in
Fig.~\protect\ref{Accel-Decel-Loop}, or it may start as a
burst-like event which then goes over into the periodic process,
as shown in the top-left insert. In the burst-like event a large
number of vortex lines is formed simultaneously at a bulk-liquid
texture instability. Here a first order transition occurs in the
order-parameter texture and it is suddenly transformed to a new
configuration in which the critical velocity is generally lower
than in the original texture. In the transformed texture the
periodic process turns on during a further increase of $\Omega$
and the corresponding $\Omega_c$ is obtained by extrapolating back
to zero peak height of the vortex satellite. Vortex formation at a
texture instability is possible only at high temperatures ($T
\gtrsim 0.7\,T_c$) where the energy barriers separating different
textures are small. The inset on the top right shows as a
histogramme the number  of vortex lines $\Delta N$ which are
produced in the burst, normalized to the equilibrium number
$\Delta N_{max}$ at the rotation velocity $\Omega_b$ after the
burst: $\Delta N_{max} = (2\pi R^2
\Omega_b/\kappa)\,(1-\beta/\sqrt{\Omega_b})$. The data for the
periodic process are marked as $(\circ)$ while burst-like events
are denoted as $(\bullet)$. Measuring conditions: ${\bf H}
\parallel \mathbf{\Omega}$, $H=$9.9 -- 15.8 mT, $P=$29.3 -- 34.2
bar. (From Ref.~\cite{RuutuA})} \label{CritVel-TextTransition}
\end{figure}

\begin{figure}[t!!!]
\begin{center}
\includegraphics[width=.7\textwidth]{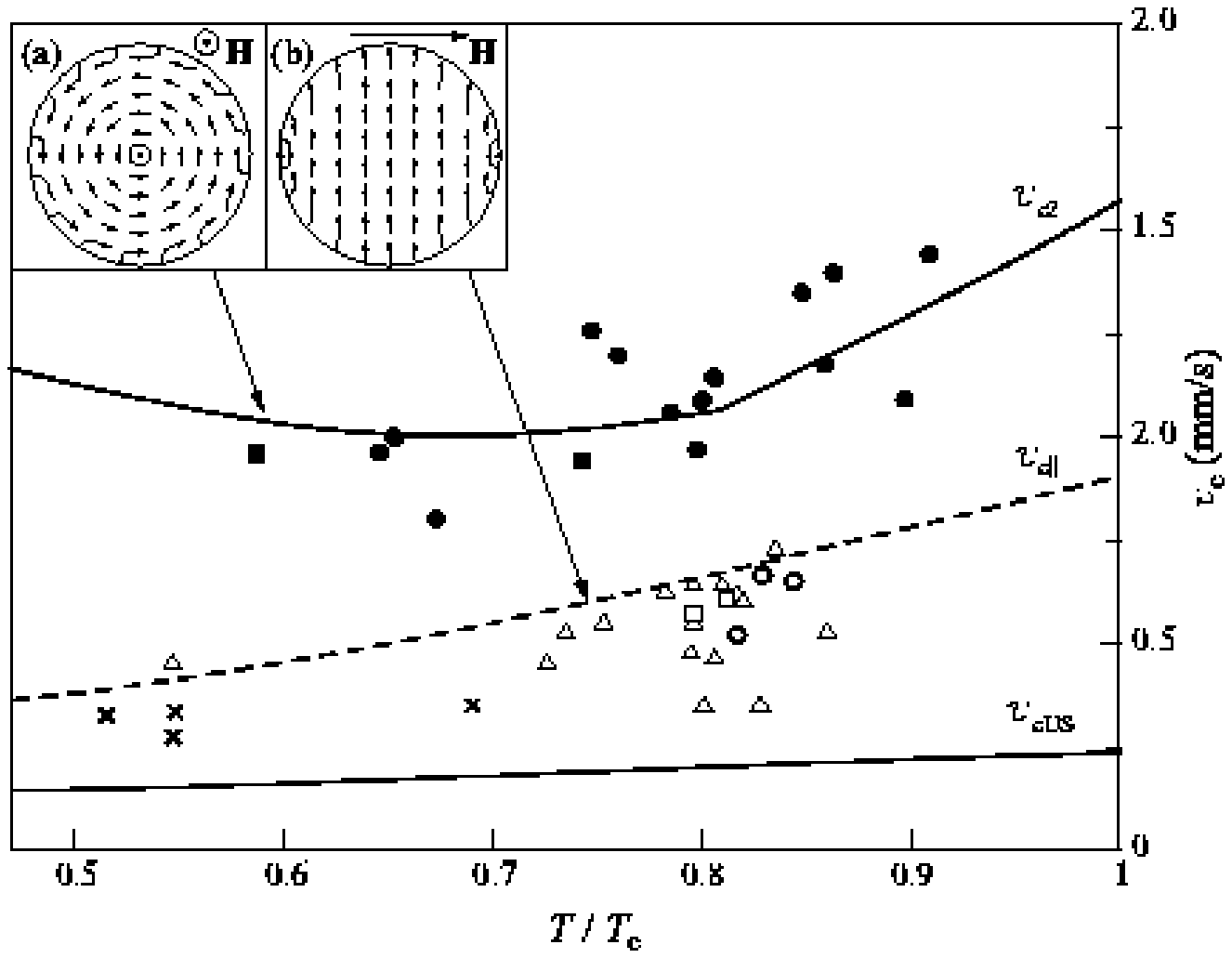}
\end{center}
\caption[Texture-CritVel]{Critical velocity of vortex formation
for selected order-parameter textures. The curves represent the
respective calculated bulk superfluid flow instability
\cite{Kopu}. The absolute instability limit at $v_{c2}$ applies to
the originally homogeneous ($\hat\mathbf{\ell}\parallel {\bf v}$)
orbital texture in inset (a). Here $\hat\mathbf{\ell}$ is confined
to the transverse plane and is dipole-locked  everywhere, when
$\Omega<\Omega_c$, except within the surface layer and in the very
center (where there is either a singular disclination line or a
dipole-unlocked radial $2\pi$ Mermin-Ho vortex). This limit can be
compared to measurements ($\bullet,\blacksquare$) in which the
sample is cooled in rotation through $T_c$, to obtain the global
equilibrium texture. The lowest $v_c$ is measured ($\times$) when
the NMR spectrum shows the signature from the transverse twist
soliton (cf. Fig.~\protect\ref{NMRspectroscopy}). These data
points can be compared to the calculated $v_c$ of a
dipole-unlocked soliton $v_{cUS}$. In transverse magnetic field,
${\bf H} \perp \mathbf{\Omega}$, the equilibrium global texture in
inset (b) is also obtained by cooling through $T_c$ in rotation
and gives data points ($\circ,\Box$) which should be compared to
the calculated instability at $v_{c \parallel}$. This texture
includes two singular disclination lines, located diametrically
opposite each other at the cylinder wall (marked with a black dot
on the horizontal diameter in inset (b)). The transverse field
measurements seem to be less sensitive to the annealing
requirement since other data ($\triangle$) with varied
prehistories are not very different. (From Ref.~\cite{RuutuA}) }
\label{Texture-CritVel}
\end{figure}

The intercept of the acceleration record with the horizontal axis,
when $\Omega \geq \Omega_c$, determines the critical velocity
$v_c$ in Fig.~\ref{Accel-Decel-Loop}. The fact that this section
is linear proves that $v_c$ remains constant during all of the
acceleration. On plotting the corresponding counterflow velocity
at the cylinder wall, $v(\Omega) = \Omega R - \nu \kappa N/(2\pi
R)$, it is seen that the noise in $v_c$ can be explained to arise
from experimental sources. Thus vortex formation proceeds here in
the form of a regular periodic process and displays no measurable
stochastic behaviour, which could be associated with nucleation
across an energy barrier.  Indeed, a simple argument shows that
the nucleation energy barrier is so large compared to thermal
energy that it cannot be overcome by any usual nucleation
mechanisms. Instead, the applied counterflow has to be increased
to the point where the barrier height goes to zero \cite{RuutuB}.
Thereby vortex formation becomes essentially an instability.

The reason for the high nucleation barrier at low applied flow
velocities  is the large length scale $\xi_D$ on which the vortex
has to be formed. The energy stored per unit length in the
superflow around the vortex core is of order $\sim \rho_s \nu^2
\kappa^2$. This has to be compared to the kinetic energy of the
applied superflow at the cylinder wall in a volume comparable to
that where the instability occurs, $\sim \rho_s v^2 \xi_D^2$. Thus
the flow velocity for creating the vortex has to be of order $v_c
\sim \nu \kappa/(2\pi \xi_D)$. This shows that the velocity for
reaching the instability decreases with length scale and is in
$^3$He-A comparable to the dipole velocity $v_D \propto 1/\xi_D$,
ie. the velocity required to break dipole locking. The barrier
height, in contrast, increases $\propto \xi_D^2$. In $^4$He-II the
appropriate length scale is the superfluid coherence length $\xi
\sim 0.1$ nm, which is of atomic size, and gives a barrier height
of order 1 K. In $^3$He superfluids the barrier is higher and the
temperature lower, both by at least three orders of magnitude.
Therefore vortex-formation takes place at an instability.

As outlined above, one might think that the instability-determined
critical velocity in $^3$He-A is a well-defined quantity. It
should depend only on the externally applied conditions, such as
temperature $(T)$ and pressure $(P)$, which determine the $^3$He-A
properties $\rho_s$ and $\xi_D$. In a smooth-walled container it
should not depend on the wall properties, since in the A phase
vortex formation has to occur essentially within the bulk liquid:
At the cylinder wall $\hat {\bf \ell}$ is oriented perpendicular
within a surface layer of width comparable to $\xi_D$, owing to a
rigid boundary condition. However, experimentally it is
immediately concluded that the formation process has a lot of
variability: It can have the regular appearance shown in
Fig.~\ref{Accel-Decel-Loop}, or it can take the burst-like form
shown in Fig.~\ref{CritVel-TextTransition}. Also the magnitude of
$v_c$ depends on the previous experimental history which the
sample has been subjected to in the superfluid state
\cite{RuutuA}.

All this variation in the characteristics of the critical velocity
can be measured in one and the same smooth-walled sample
container. This is quite different from $^3$He-B in the same
cylinder, where vortex formation proceeds as a reproducible and
well-behaved regular process of similar form as in
Fig.~\ref{Accel-Decel-Loop} \cite{RuutuB}. In $^3$He-B the
critical velocity is typically at least an order of magnitude
larger, as can be expected when a singular-core vortex has to be
formed on the length scale of the superfluid coherence length
$\xi(T,P) \sim$ 10 -- 100 nm.

This seemingly unruly behaviour of $^3$He-A can be explained by
the dependence of the instability velocity on the global order
parameter texture, which is still a poorly controlled and
understood feature of the experiments. In
Fig.~\ref{CritVel-TextTransition} various measurements on $v_c$
have been collected, regardless of the earlier history of the
samples. The data points, accumulated from three different sample
cylinders, seem to fall between a maximum and a minimum limit. In
particular, there are no measurements with zero or very small
$v_c$. The global order parameter texture in the cylinder depends
on the history of sample preparation. In
Fig.~\ref{Texture-CritVel} the measurements are grouped according
to what type of global texture is expected on the basis of the
procedure which was used to prepare the sample. This figure now
provides some credibility to the notion that the global texture
can be influenced by the sample preparation method, that the
critical velocity indeed depends on the texture, and that
theoretically calculated estimates \cite{Kopu} of the bulk-liquid
flow instability for the different textures provide reasonable
upper or lower bounds for the measured data. Thus the measurement
of $v_c$ seems to provide the first experimental tool for
characterizing the global texture, however indirectly. Also the
independence of the observed features on the container exemplifies
the fact that in the A phase vortex formation is a truly intrinsic
bulk-liquid process, well separated from the container wall
(unlike even $^3$He-B).

The lowest critical velocities in Fig.~\ref{Texture-CritVel} are
recorded when a dipole-unlocked soliton (cf.
Fig.~\protect\ref{NMRspectroscopy}) is present in the container
and is oriented perpendicular to the rotation axis.  The soliton
is a planar wall of width $\sim \xi_D$ which separates bulk liquid
in two different, but degenerate minima of the dipole energy,
$\hat{\mathbf{d}} \uparrow\uparrow \hat\ell$ and $\hat{\mathbf{d}}
\uparrow\downarrow \hat\ell$, while within the wall the dipole
energy is not minimized. This explains the reduced critical
velocity since in this case a dipole-unlocked region exists at the
cylinder wall where the counterflow velocity is maximized and
which can seed the formation of a dipole-unlocked vortex. The
resulting structure, the intersection of a double-quantum vortex
line with a transverse soliton sheet, is an example of a
metastable $\hat \mathbf{\ell}$ field with complicated knot-like
continuous topology. This unstable configuration can be maintained
in a long cylinder to moderately high rotation $(\Omega \lesssim
0.5$ rad/s) \cite{VorSolIntersection}.

However, the most surprising case is that when a vertical
dipole-unlocked soliton is present. Here the soliton sheet is
oriented approximately parallel to the rotation axis and connected
to the cylinder wall along two dipole-unlocked lines over the
whole length of the container. As a function of $\Omega$ the
critical velocity is then initially close to zero and increases in
a nonlinear fashion. In rotation this situation leads to a new
stable order-parameter structure which is known as the {\it vortex
sheet}, a combined topologically stable object with planar
structure into which linear quantized circulation has been
integrated.

\section{Vortex sheet}

\begin{figure}[t!!!]
\begin{center}
\includegraphics[width=1.0\textwidth]{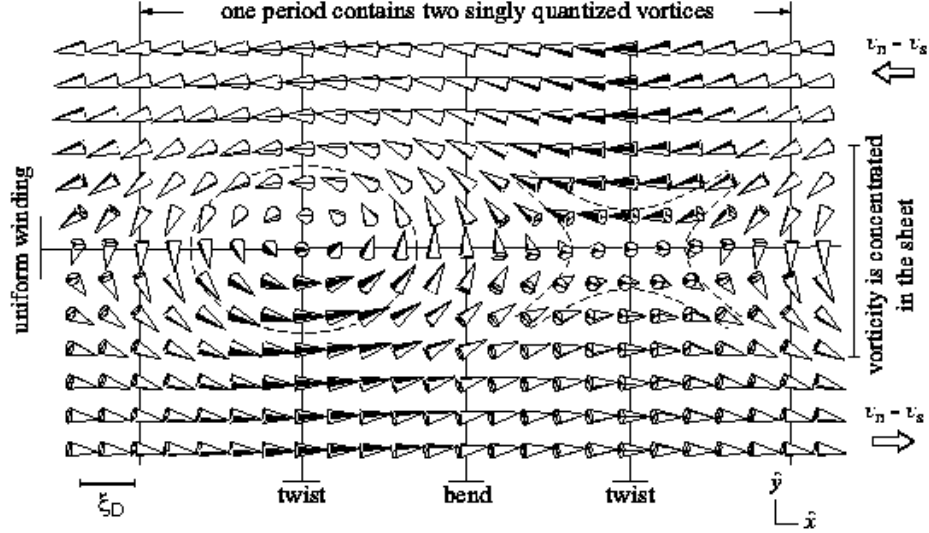}
\end{center}
\caption[VorSheet-OrderParam]{Orbital $\hat\mathbf{\ell}$ field of
the vortex sheet. Outside the sheet $\hat\mathbf{\ell}$ is
oriented parallel to the sheet, but in opposite directions on the
two sides of the sheet. The sheet itself consists of a periodic
linear chain of alternating circular and hyperbolic Mermin-Ho
vortices, the two constituents of the double-quantum vortex in
Fig.~\protect\ref{CUV}. (From Ref.~\cite{Lexicon}) }
\label{VorSheet-OrderParam}
\end{figure}

\begin{figure}[t!!!]
\begin{center}
\includegraphics[width=.7\textwidth]{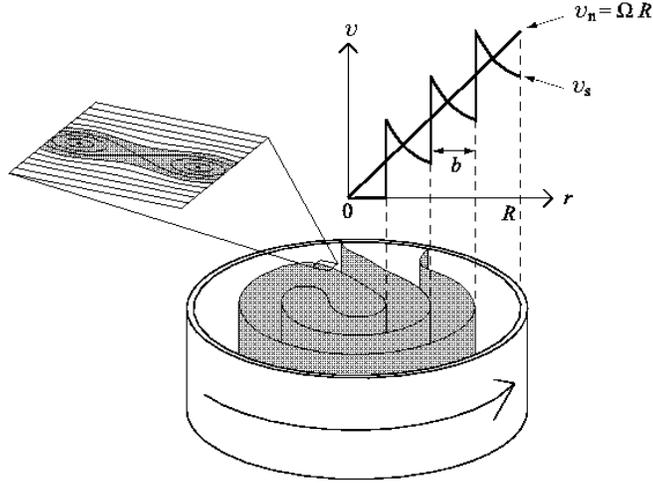}
\end{center}
\caption[VorSheet-LargeScale]{Large-scale structure of the vortex
sheet in its equilibrium configuration in the rotating container.
The vortex sheet is formed when $^3$He-A is accelerated to
rotation in an applied magnetic field $(H>H_D)$ and a soliton wall
parallel to the rotation axis exists in the container. If the
magnetic field is oriented along $\mathbf{\Omega}$, the
equilibrium configuration of the folding is a double spiral. The
graph on the top right illustrates the radial distributions of the
normal and superfluid velocities. The latter is discontinuous
across the sheet and differs from that of individual vortex lines
in Fig.~\protect\ref{RotBucket}. } \label{VorSheet-LargeScale}
\end{figure}

In classical turbulence a vortex sheet is known as a thin
interface across which the tangential component of the flow
velocity is discontinuous and within which the vorticity
approaches infinity \cite{Saffman}. In superfluids vortex sheets
with quantized circulation, which separate two regions with
irrotational flow, were briefly discussed in the late 1940's and
early 1950's, to explain the observations from rotating
experiments \cite{Donnelly}. However, in an isotropic superfluid
like $^4$He-II the vortex sheet was not found to be stable with
respect to break up into isolated vortex lines. In quantum systems
the vortex sheet was first experimentally identified in the
anisotropic $^3$He-A phase in 1994 \cite{Parts-VorSheet}. The
existence of vortex sheets in the form of domain walls which
incorporate half-integer magnetic flux quantization is also
discussed in unconventional superconductors with $^3$He-A-like
structure such as Sr$_2$RuO$_4$, UPt$_3$ or U$_{\rm
{1-x}}$Th$_{\rm x}$Be$_{13}$ \cite{Sigrist}.

In $^3$He-A the vortex sheet is formed from a soliton wall into
which Mermin-Ho vortices with continuous structure have been
incorporated. As shown in Fig.~\ref{VorSheet-OrderParam}, the
sheet is made up of a alternating linear chain of singly quantized
vortices with circular and hyperbolic $\hat\mathbf{\ell}$ winding
such that a fully periodic structure results. Like in the case of
vortex lines, the vortex sheet is translationally invariant in the
direction parallel to the rotation axis $\mathbf{\Omega}$. The
large-scale structure in the equilibrium configuration is a
continuously meandering foil which is attached at both ends at two
vertical connection lines to the cylindrical wall
(Fig.~\ref{VorSheet-LargeScale}). It is these connection lines
where circulation quanta are added to or removed from the sheet.

The hydrodynamic stability of the vortex sheet was calculated by
Landau and Lifshitz \cite{Landau}. By considering the equilibrium
state with the kinetic energy from the flow between the folds and
the surface tension $\sigma$  from the soliton sheet, it is
concluded that the distance between the parallel folds has to be
$b = (3 \sigma/ \rho_{s \parallel})^{1/3} \; \Omega^{-2/3}$. This
is somewhat larger than the inter-vortex distance in a cluster of
vortex lines. The areal density of circulation quanta has
approximately the solid-body value $n_v = 2\Omega/\kappa$. This
means that the length of the vortex sheet per two circulation
quanta is $p = \kappa/(b\Omega)$, which is the periodicity of the
order-parameter structure in Fig.~\ref{VorSheet-OrderParam}. The
NMR absorption in the vortex-sheet satellite measures the total
volume of the sheet which is proportional to $1/b \propto
\Omega^{2/3}$. The nonlinear dependence of absorption on rotation
is the experimental signature of the vortex sheet, in addition to
its low value of critical velocity.

\begin{figure}[t!!!]
\begin{center}
\includegraphics[width=.95\textwidth]{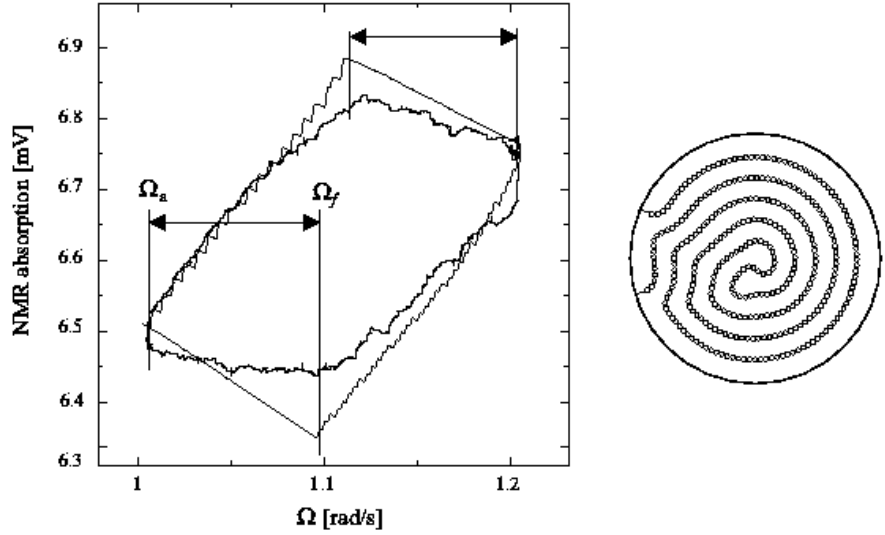}
\end{center}
\caption[CritVel-VorSheet-Measurement]{Response of the equilibrium
vortex sheet to a closed acceleration -- deceleration cycle in the
rotation drive. This measurement is used to define the critical
velocity of the vortex sheet as a function of $\Omega$. During
increasing $\Omega$ the threshold, where more circulation is
added, is denoted as $\Omega_f$, while the threshold of
annihilation, where circulation is removed,  is $\Omega_a$. The
experimental result (thick line) is compared to a simulation
calculation \cite{Simulation} (thin line) on the vortex-sheet
configuration which is shown on the right at $\Omega_a = 1.0$
rad/s. In the vortex-sheet meander the circles denote the actual
center positions of each circulation quantum. The measurement has
been performed in a smooth-walled fused quartz cylinder. The
measuring conditions are the same as in
Figs.~\protect\ref{VorSheetHarmonicResponse} and
\ref{VorSheetStepResponse}. } \label{CritVel-VorSheet-Measurement}
\end{figure}

\begin{figure}[t!!!]
\begin{center}
\includegraphics[width=.7\textwidth]{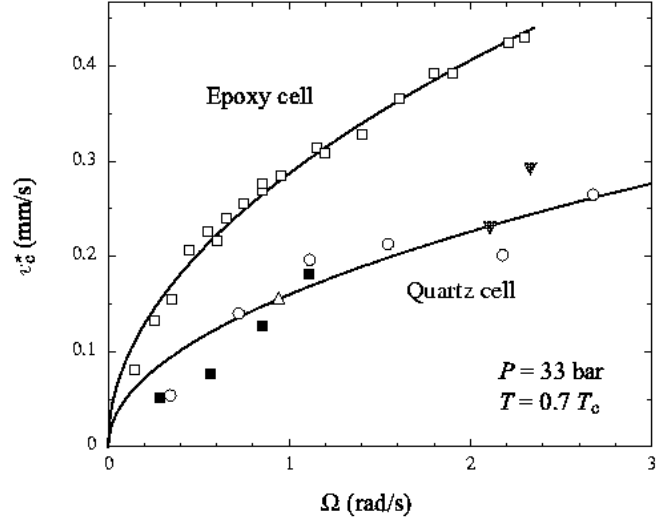}
\end{center}
\caption[CritVel-VorSheet]{Measured critical velocity $v_c^*$ of
vortex-sheet expansion, when new circulation is added as a
function of the applied rotation $\Omega$. In contrast to the
regular periodic process in vortex-line formation, where $v_c$ is
a constant as a function of $\Omega$, here $v_c^*(\Omega) \propto
\sqrt{\Omega}$. The measurements represent the situation when a
vertical soliton wall is already present in the cylinder and the
vortex sheet has been adiabatically grown in an axially oriented
magnetic field to its equilibrium configuration. The solid curves
are fits with $\sqrt{\Omega}$ dependence. The rougher epoxy wall
has a larger magnitude of critical velocity, perhaps because of
pinning of the connection lines which resists smooth readjustment
in the folding of the vortex sheet as a function of $\Omega$. The
different symbols of data points (quartz cell) illustrate the
reproducibility of the results from one adiabatically grown vortex
sheet to another. } \label{CritVel-VorSheet}
\end{figure}

The vortex sheet is formed whenever a vertical dipole-unlocked
soliton sheet is present in the container and rotation is started.
The reason for this is the low critical velocity at the
dipole-unlocked connection lines between the sheet and the
cylinder wall. This facilitates the creation of new circulation
quanta and lowers the critical velocity in the sheet well below
that of isolated vortex lines. When a new circulation quantum is
added to the sheet, it experiences repulsion from the circulation
which already resides in the sheet close to the connection line.
Owing to this additional $\Omega$-dependent barrier the critical
velocity becomes $\Omega$-dependent. Experimentally the critical
velocity is defined from Fig.~\ref{CritVel-VorSheet-Measurement}
as $v^*_c = (\Omega_f - \Omega_a)R$, the separation in rotation
velocities between the thresholds where new circulation is formed
$(\Omega_f)$ and existing circulation is annihilated $(\Omega_a)$.

The measured critical velocity in Fig.~\ref{CritVel-VorSheet}
follows qualitatively the dependence $v^*_c(\Omega) \propto
\sqrt{\Omega}$. This approximate relation illustrates the
characteristics of the vortex sheet: The Magnus force $F_M =
\rho_s \kappa v$ from the vortex-free counterflow, with the
velocity $v = (\Omega - \Omega_a)R \approx 2d \Omega$ at the
cylinder wall, attempts to pull additional circulation into the
sheet. It is opposed by the repulsion from the circulation, which
already resides in the sheet closest to the connection line, at a
distance equal to the width $d$ of the circulation-free annulus.
The balance between the repulsion $F_r = (\rho_s \kappa^2 /d)
\ln{[d/ (\alpha \xi_D)]}$ and $F_M$ gives a critical counterflow
velocity of order $v_c \sim \sqrt{2\kappa\Omega}$.

Actually, because of the presence of the two connection lines, a
vortex-sheet state does not have full axial symmetry in the
distribution of its vortex-free counterflow even in the
double-spiral configuration of
Fig.~\ref{CritVel-VorSheet-Measurement} (ie. at the connection
line the distance of the first circulation quantum from the wall
is different from that in the outermost circular spiral part).
This causes a small difference in the measured $v^*_c =(\Omega_f -
\Omega_a)R$ from the real value of $v_c$ at the connection lines.
The two can be connected if the sheet configuration is known. A
qualitative difference between these two critical velocities is
that the true value of $v_c$ does not approach zero when $\Omega
\rightarrow 0$, but remains finite. This is expected since some
counterflow velocity is required when new circulation is formed
even at a dipole unlocked connection line of a circulation-free
soliton sheet, since also the attractive interaction of the
emerging vortex line with its image forces within the wall have to
be overcome.

\section{Dynamic response}

\begin{figure}[t!!!]
\begin{center}
\includegraphics[width=.73\textwidth]{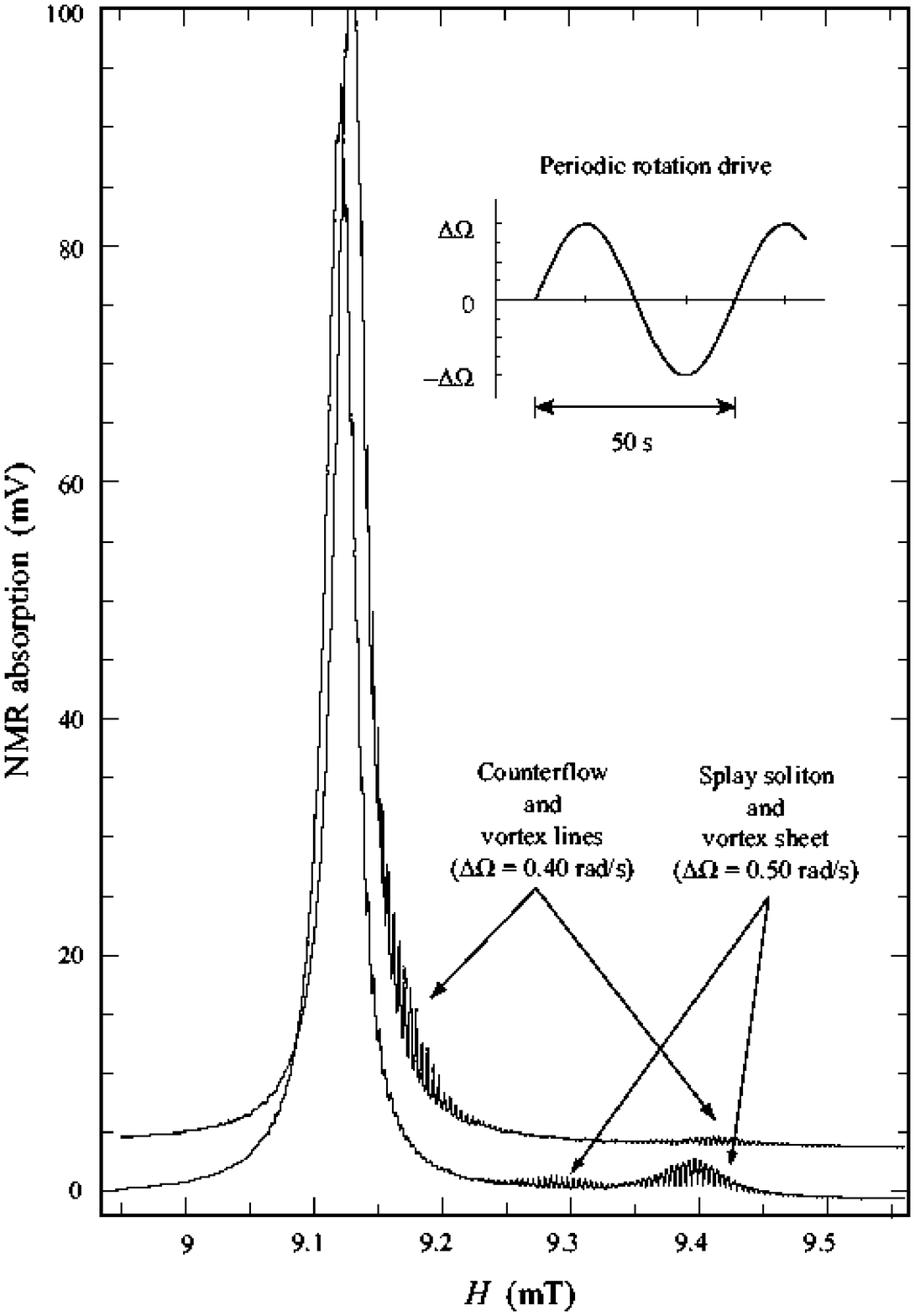}
\end{center}
\caption[VorSheet-Confinement]{NMR absorption spectra in
back-and-forth rotation. The measurement is performed by sweeping
slowly the polarizing magnetic field $H$ at constant rf excitation
frequency $(f=310$ kHz) across the NMR region while the sample is
in sinusoidal back-and-forth rotation: $\Omega(t) = \Delta \Omega
\, \sin{\omega t}$. The field sweep is much slower than the period
of the rotation drive ($2\pi/\omega = 50$ s). Thus the rotation
appears as a modulation envelope on the field-dependent NMR
absorption. 1) When the amplitude $\Delta \Omega $ of the
sinusoidal rotation is sufficiently small $( v_D/R \lesssim \Delta
\Omega = 0.40$ rad/s) the circulation enters in the form of
double-quantum vortex lines (\emph{upper spectrum}). Here the
absorption is transferred periodically between the vortex and
counterflow satellites (spectrum {\bf CUV + CF} in
Fig.~\protect\ref{NMRspectroscopy}). 2) When the rotation
amplitude is increased $(\Delta \Omega = 0.50$ rad/s) the
circulation goes into the vortex sheet \emph{(lower spectrum)}.
Now the absorption is shifted between the vortex-sheet and soliton
satellites (spectra labeled as {\bf VS} and {\bf SS} in
Fig.~\protect\ref{NMRspectroscopy}). (Measuring conditions:
$2R=3.87$ mm, ${\bf H} \parallel \mathbf{\Omega}$, $P=34$ bar,
$T=0.7\,T_c$) } \label{VorSheet-Confinement}
\end{figure}

\begin{figure}[t!!!]
\begin{center}
\includegraphics[width=0.90\textwidth]{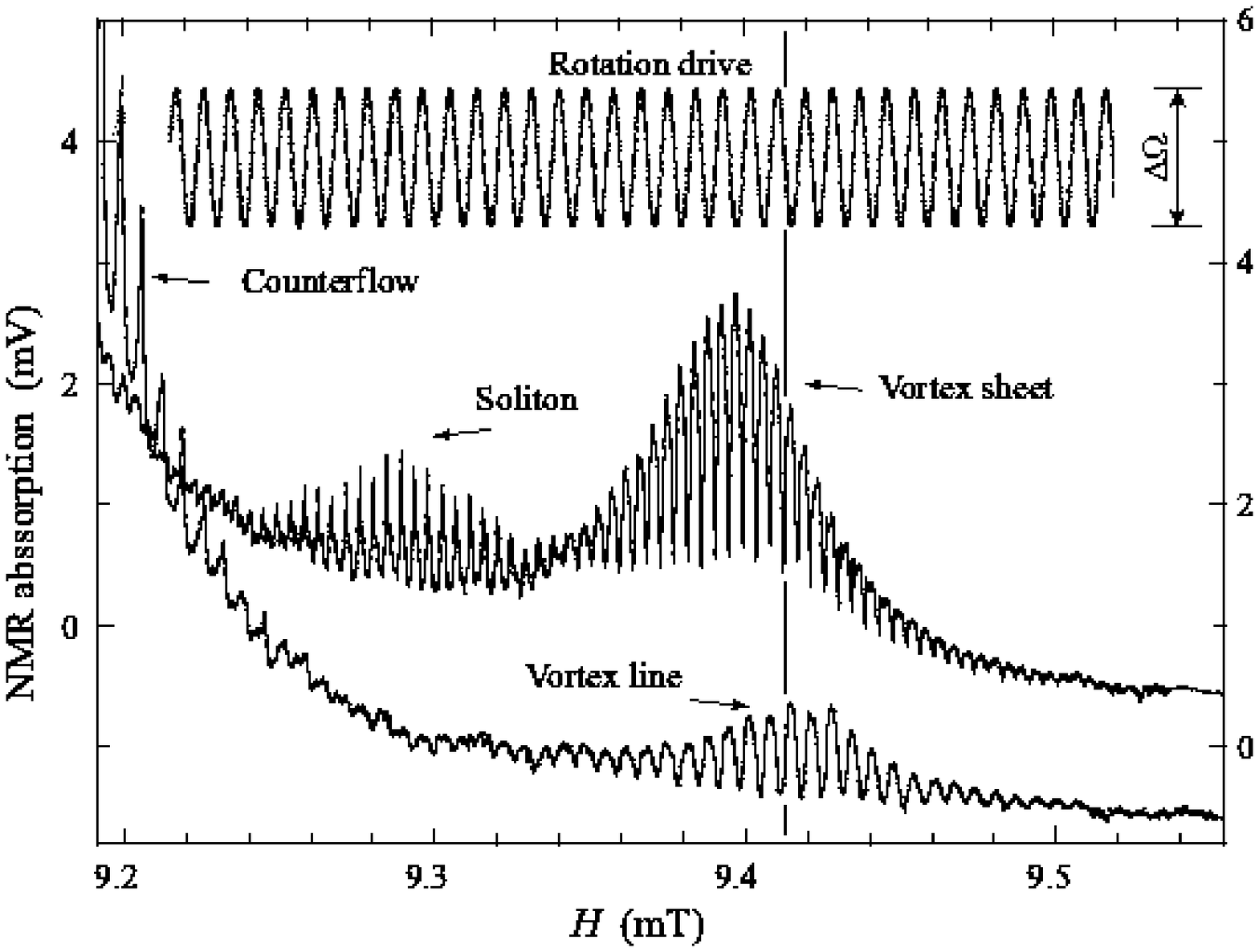}
\end{center}
\caption[VorSheetSatModulation]{Satellite absorptions from
Fig.~\protect\ref{VorSheet-Confinement}. The topmost trace shows
the correct phase of the rotation drive with respect to that of
the NMR absorptions in the different satellites. The horizontal
axis is actually time $t$ since all variables, the NMR field sweep
$H(t)$, rotation $\Omega (t)$, and the two absorption signals,
have been plotted on a common time axis. }
\label{VorSheetSatModulation}
\end{figure}

So far we have looked at the response of the $^3$He-A
order-parameter field to essentially adiabatic changes in the
rotation velocity. But what happens if the rotation is changing
rapidly with time? Because of its lower critical velocity, it is
found that the vortex sheet becomes the preferred structure in
rapidly changing rotation, rather than individual double-quantum
vortex lines. Consequently in $^3$He-A, the response to
high-frequency perturbations in the rotation drive is explained by
the dynamic properties of the vortex sheet. The central feature
becomes the interplay between the large-scale configuration of the
sheet and its confined circulation. The first illustration of
these considerations is provided by the measurements in
Fig.~\ref{VorSheet-Confinement}.

In the experiment of Fig.~\ref{VorSheet-Confinement} the complete
NMR absorption spectrum is slowly recorded while the rotation is
harmonically swinging back and forth around zero with a short
period of 50 s. At low amplitude of back-and-forth rotation the
circulation enters in the form of double-quantum vortex lines
(upper spectrum). The lines are periodically formed and then
annihilated during both the positive and negative half cycles of
rotation. During the positive and negative half cycles the
vorticity has opposite sign of circulation: All vortex lines are
annihilated approximately by the time when rotation goes through
zero. Therefore the envelope of the vortex-line satellite is
modulated at twice the frequency of the rotation drive. To record
the spectrum the rate of the NMR field sweep has to be much slower
than the period of the rotation drive: The horizontal scale, which
here is plotted in terms of the linearly changing NMR field sweep,
is actually the common time axis for all variables.

A rarely seen feature of the upper spectrum is the modulation in
the absorption on the high-field flank of the large bulk-liquid
peak. This absorption component is created by the slight dipole
unlocking in the vicinity of the surface layer on the cylinder
wall when the counterflow velocity gets sufficiently large. The
modulation of this absorption is approximately in anti-phase with
that in the vortex-line satellite, ie. minimum absorption in the
vortex-satellite is reached approximately at the same time when
absorption in the counterflow signal is at maximum. Since the
total area in the absorption spectrum must be constant as a
function of time, this means that the absorption is transferred
periodically between the vortex and counterflow satellites.

The lower spectrum in Fig.~\ref{VorSheet-Confinement} is recorded
with a larger amplitude of back-and-forth rotation. In this case
the circulation enters in the form of a vortex sheet, which now
provides a more stable response than double-quantum vortex lines.
(During the first few half cycles the response may be in the form
of lines, but it soon goes spontaneously over into the vortex
sheet.) The modulation of the absorption in the vortex-sheet
satellite is similar to that of the double-quantum vortex peak.
The major change  is the absence of the counterflow signal, since
now the critical velocity is lower and the counterflow velocity is
limited below a smaller value. The second major difference is the
presence of the absorption in the splay soliton satellite. This
absorption component appears when the circulation is absent around
$\Omega \approx 0$ and the vortex-sheet satellite is at minimum.
Thus in the lower spectrum the absorption is transferred between
the vortex-sheet and soliton satellites and their modulation
signals are in anti-phase.

The high-field tails of the spectra are repeated in larger scale
in Fig.~\ref{VorSheetSatModulation}, to display the relative
phases of the different variables. A comparison of these satellite
signals demonstrates concretely the differences between the
responses in terms of linear and planar states of vorticity. The
vortex sheet satellite is larger in amplitude because the critical
velocity is lower and the amplitude $\Delta \Omega $ of the
rotation drive larger; therefore the peak represents more
circulation quanta than that of the double-quantum vortex lines.
Also the vortex sheet absorption tracks more closely the phase of
the rotation drive (topmost signal), again because of the smaller
critical velocity. This is also the explanation for the pointed
and narrow minima in the modulation envelope of the vortex-sheet
satellite and the corresponding maxima in the soliton satellite.
The good reproducibility of the modulated soliton absorption shows
that the soliton remains stable in the container at $\Omega = 0$.
Actually, in this measurement the configuration of the vortex
sheet is not that of the equilibrium double spiral but one where
multiple pieces of sheets exist which all are attached along two
contact lines to the vertical cylinder wall \cite{Eltsov}.

\begin{figure}[t!!!]
\begin{center}
\includegraphics[width=0.85\textwidth]{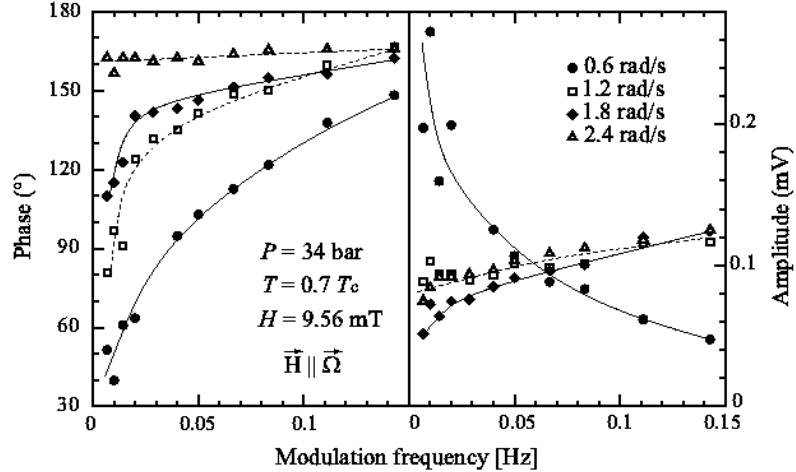}
\end{center}
\caption[VorSheetHarmonicResponse]{Dynamic response of the
equilibrium vortex sheet to harmonic modulation of the rotation
drive: $\Omega (t) = \Omega_0 + \Delta \Omega \, \sin{\omega t}$.
The left panel shows the phase of the modulated absorption
component in the peak height of the vortex-sheet satellite,
relative to the rotation drive (in degrees). The right panel gives
the amplitude of the modulated absorption (peak to peak), referred
to in mV at the output of the cryogenic preamplifier operated at
LHe temperature \cite{RuutuB}. The solid curves are guides to the
eye. The measuring conditions are the same as in
Fig.~\protect\ref{VorSheet-Confinement} with $\Delta \Omega =
0.050$ rad/s. (From Ref.~\cite{LT22}) }
\label{VorSheetHarmonicResponse}
\end{figure}
\begin{figure}[t!!!]
\begin{center}
\includegraphics[width=.75\textwidth]{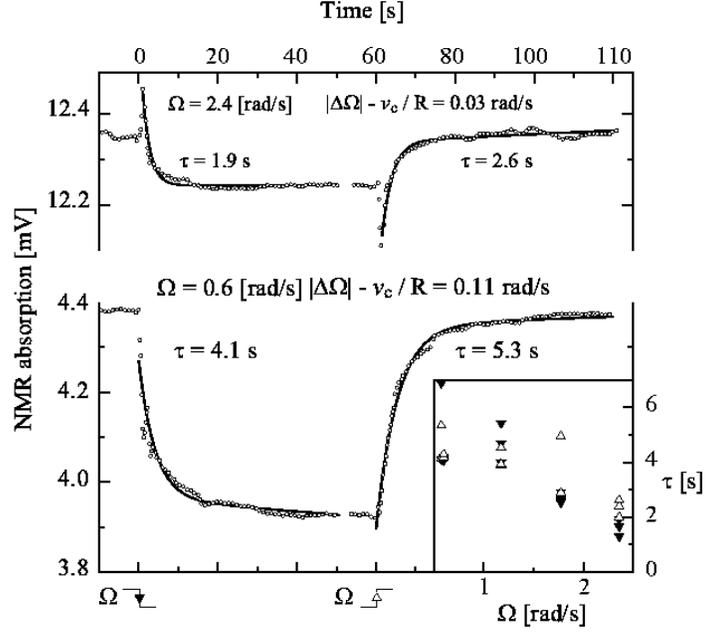}
\end{center}
\caption[VorSheetStepResponse]{Response of the equilibrium vortex
sheet to a step change in the rotation drive. On the left $\Omega$
is reduced by $\Delta \Omega = 0.15$ rad/s and on the right it is
increased back to the original value. The solid curves are
exponential fits. The corresponding time constants are plotted in
the inset, with ($\blacktriangledown$) for downward and
$(\triangle$) for upward step changes. (From Ref.~\cite{LT22}) }
\label{VorSheetStepResponse}
\end{figure}

A straightforward continuation of the measurements in
Fig.~\ref{VorSheetSatModulation} is to monitor the satellite peak
height as a function of the modulation frequency, to find the
dynamic response to changes in the rotation drive. In the case of
double-quantum vortex lines, the satellite intensity (or peak
height at constant line width) monitors in first approximation the
number of lines. In this situation only the fast radial motion of
vortex lines can be measured. This is performed by stopping
rotation very abruptly and by measuring the time dependence of
vortex-line annihilation. The tail of this signal is given by
$N(t) = N(0) (1+t/\tau_F)^{-1}$, where $N(0)$ is the number of
vortex lines in the container at the moment $t=0$, when rotation
has come to a stop. Such measurements give the characteristic
decay time $\tau_F = (\nu\kappa n_v(0)\; \rho_nB/\rho)^{-1}$,
where $\nu\kappa n_v(0)$ is the density of circulation quanta at
$t=0$ and $\rho_nB/\rho$ is the dissipative mutual friction. This
mode of vortex-line motion has been monitored in the $^3$He
superfluids for double-quantum vortices in A phase \cite{Hakonen}
and for singular-core vortices in B phase \cite{Kondo}. Both
measurements are consistent with other determinations of mutual
friction \cite{Hook}. Other resonance techniques can be used to
record also the slow approach of vortex lines to equilibrium at
constant $\Omega$, after an initial disturbance has been switched
off. This motion is predominantly azimuthal in character for an
isolated vortex cluster and in the $^3$He superfluids it is over
damped, ie. has pure exponential time dependence \cite{Kondo}.

In the vortex sheet the circulation is topologically confined
within the sheet where it can move in two ways: either along the
sheet or by forcing the entire sheet to contract or expand. Both
types of motion proceed only via changes in the length and folding
of the sheet. A slow final readjustment in the configuration of
the sheet follows these faster initial responses, if the texture
is left to anneal at constant $\Omega$. The central feature is
thus the interplay between the large-scale configuration of the
sheet and its confined circulation. This aspect is illustrated by
the measurements in Figs.~\protect\ref{VorSheetHarmonicResponse}
and \ref{VorSheetStepResponse}.

The satellite intensity of the vortex sheet is proportional to the
total volume of the sheet. The circulation is distributed as a
continuous periodic chain along the sheet where the total number
of quanta $N$ is given by the length $L$ of the meander in the
transverse plane: $N = L/(p/2) = 2bL\Omega/\kappa$. Thus the
satellite signal has to be proportional to
\begin{equation} L = {1 \over 2}\kappa N \left( {\rho_{s \parallel} \over {3 \sigma
\Omega}} \right)^{1/3} \;, \end{equation}
\label{VorSheetLength}which means that it displays a dependence on
$\Omega$ even at constant $N$, as evident in
Fig.~\ref{CritVel-VorSheet-Measurement}. Therefore, also the
readjustment to small disturbances in $\Omega$ at constant
circulation can be measured for the vortex sheet with this
technique \cite{LT22}.

The response of the peak height of the vortex-sheet satellite is
shown in Fig.~\ref{VorSheetHarmonicResponse} for harmonic
modulation of the rotation drive and in
Fig.~\ref{VorSheetStepResponse} for a step change. In both cases
the change by $\Delta \Omega $ about the average rotation
$\Omega_0$ is kept as small as possible, to maintain $N$ constant.
However, because of finite measuring resolution the condition
$2\Delta \Omega < \Omega_f - \Omega_a$  (cf.
Fig.~\ref{CritVel-VorSheet-Measurement}) is valid only at large
$\Omega_0$ in the measurement of the harmonic response. In the
step-response measurement it is not satisfied even at the highest
$\Omega_0$ value.

Let us first look at the characteristics in the limit of large
$\Omega_0$. The left panel in Fig.~\ref{VorSheetHarmonicResponse}
shows that the phase shift between the response and the drive is
then approaching $180^{\circ}$. The right panel shows that this
response occurs at small amplitude. The out-of-phase behaviour at
constant $N$ is expected from Eq.~(\ref{VorSheetLength}): To
maintain solid-body-rotation, the vortex sheet contracts during
increasing rotation and expands during decreasing $\Omega$.

This out-of-phase signal at constant $N$ identifies the origin of
the sharp out-of-phase spike in the step response at large
$\Omega_0$ (upper trace in Fig.~\ref{VorSheetStepResponse}). The
spike is then followed by a slow exponential in-phase response
which has to be associated with the change in $N$. The amplitude
of this component is small at large $\Omega_0$ (upper trace) but
grows much larger at low $\Omega_0$ where the change in $N$
becomes the dominant effect. The time constant of the exponential
signal is plotted as a function of $\Omega_0$ in the inset at
bottom right in Fig.~\ref{VorSheetStepResponse}.  In the harmonic
response this means that with decreasing $\Omega_0$ the phase
shift is reduced and the amplitude grows.

The time sequence of different processes in the dynamic response
of the equilibrium vortex sheet with double-spiral configuration
is thus the following: (i) The fastest response with a time
constant of order 1\,s occurs in the density of the circulation
quanta. In the axial field with circularly spiraling folding this
can happen only perpendicular to the sheet, ie. the sheet either
contracts or expands at constant $N$. (ii) The adjustment of the
number of circulation quanta to the equilibrium value occurs
slower on a time scale of order 10\,s. This requires the motion of
the circulation along the sheet in unison and the corresponding
readjustment in the length of the sheet. (iii) The slowest
component is the annealing of the vortex sheet at constant
$\Omega$ which takes place on a time scale of minutes. In the
present range of high and intermediate temperatures all motions
are exponentially over damped -- a generic property of $^3$He
superfluids where the kinematic viscosity of the normal component
is so large that it can be considered to be clamped to corotation
with the container.

More recent measurements demonstrate that the response of $^3$He-A
to a rotation drive, which includes components at high frequency
and large amplitude, is dominated by the vortex sheet and its
properties. Again the basic features are here derived from
structural considerations: The large-scale configuration of the
vortex sheet is altered substantially from the equilibrium state
such that the dynamic response becomes greatly enhanced
\cite{Eltsov}. This leads to considerable gains in the kinetic
energy under rapidly changing flow conditions which readily 
compensate the small increases in textural energies.

\section{Summary and future work}

The hydrodynamics of $^3$He-A will undoubtedly be a source for
more surprises. It is well known that with heat flow superfluid --
normal fluid thermal counter currents can be produced which drive
dissipative time-dependent $\hat \mathbf{\ell}$ textures
\cite{HeatFlow}. In rotational flow a large number of different
structures of quantized vorticity have been identified. These make
it possible for the superfluid to mimic optimally solid-body
rotation, depending on temperature, magnetic field, the
oder-parameter texture, and the properties of the rotation drive.

NMR has been the most efficient method for distinguishing between
different vortex textures, by providing the possibility to probe
the order-parameter texture from outside the rotating container.
The method is well suited for the measurement of a vortex cluster
which is translationally invariant in the direction parallel to
the rotation axis. However, both the vortex texture and the NMR
spectrum change if the orientation of the polarization field is
rotated. Therefore the NMR method is less suited for precise
measurements of vortex states which are less regular in
configuration than rotating vortex clusters. An exception are
random tangles of vorticity where the \emph{average density} of
dipole-unlocked  vorticity can be continuously monitored.

Uniform rotation is not generally suited for the study of
turbulence. Sufficiently long-lived turbulent states, which could
efficiently be recorded with present continuous-wave NMR
techniques, have not been produced by rotating $^3$He-A at
temperatures above $0.5\, T_c$. At lower temperatures the normal
fluid fraction is rapidly depleted and the hydrodynamic response
time of the superfluid fraction increases, owing to the reduced
coupling with the walls of the rotating container. If the A phase
can be maintained in metastable state in low magnetic fields to
these low temperatures \cite{Schiffer}, then it might become
possible to generate transient, but long-lived turbulent states by
rotation techniques. For such studies present measuring techniques
would be adequate. We might thus expect to see new development in
the hydrodynamics of $^3$He-A which would illuminate the
properties of an anisotropic superfluid in the zero temperature
limit.

\vspace{5mm}

\emph{Acknowledgements:}  This work was funded in part by the EU
-- Improving Human Potential -- Access to Research Infrastructures
programme under contract EC HPRI-CT-1999-50.


%

\end{document}